\documentclass[twocolumn,aps,prl,amssymb,showpacs,groupedaddress,superscriptaddress]{revtex4-1}
\usepackage{amsfonts}
\usepackage{graphicx}
\usepackage{subfigure}%
\usepackage{bm}%

\begin{document}


\title{The Pairing of Spin-orbit Coupled Fermi Gas in Optical Lattice}


\author{Ho-Kin Tang}
\affiliation{Beijing Computational Science Research Center,Beijing, 100084, China}
\affiliation{Department of Physics, The Chinese University of Hong Kong, Hong Kong, China}

\author{Xiaosen Yang}
 \email{yangxs@csrc.ac.cn}
\affiliation{Beijing Computational Science Research Center,Beijing, 100084, China}

\author{Jinhua Sun}
\affiliation{Beijing Computational Science Research Center,Beijing, 100084, China}

\author{Hai-Qing Lin}
 \email{haiqing0@csrc.ac.cn}
\affiliation{Beijing Computational Science Research Center,Beijing, 100084, China}



\date{\today}

\begin{abstract}
We investigate Rashba spin-orbit coupled Fermi gases in square optical lattice
by using the determinant quantum Monte Carlo (DQMC) simulations which is free
of the sign-problem. We show that the Berezinskii-Kosterlitz-Thoules phase
transition temperature is firstly enhanced and then suppressed by spin-orbit
coupling in the strong attraction region. In the intermediate attraction
region, spin-orbit coupling always suppresses the transition temperature. We
also show that the spin susceptibility becomes anisotropic and retains finite
values at zero temperature.
\end{abstract}


\maketitle

\textit{Introduction:} Spin-orbit coupling (SOC), breaking the inversion
symmetry, has attracted extensive attentions in condensed
matter\cite{RevModPhys.82.3045,RevModPhys.83.1057}. Recently, SOC in both the
bosonic\cite{PhysRevLett.102.130401, naturelin2009synthetic} and
fermionic\cite{PhysRevLett.109.095301, PhysRevLett.109.095302} systems has been
realized in ultracold atomic experiments. These milestone breakthroughs have
opened up an exciting route to study the novel phases
\cite{PhysRevLett.97.240401,PhysRevLett.103.035301,PhysRevA.84.031608,PhysRevA.87.063610,PhysRevLett.109.085302,PhysRevLett.108.225301,PhysRevLett.105.160403,PhysRevLett.107.195305,reviewHzhai}
induced by SOC in these systems.

By introducing SOC, two dimensional (2D) fermionic systems exhibit much more
rich phenomena\cite{PhysRevLett.109.085303,
PhysRevLett.109.140402,PhysRevA.85.023633,PhysRevLett.108.025301,PhysRevLett.107.195304}.
SOC can stabilize the topological nontrivial superfluid states
\cite{PhysRevLett.103.020401,PhysRevLett.109.105302,qunaturecom2013topological,zhangnaturecomm2013topological}.
Majorana zero mode exists in the vortices of these topological nontrivial
phases and plays a crucial role in topological quantum
computation\cite{PhysRevLett.98.010506}. It was found that SOC has nontrivial
effect on pairing and
superfluidity\cite{PhysRevLett.109.105302,PhysRevLett.108.145302} in
homogeneous systems. SOC enhances the pairing but suppresses the superfluidity.
On lattice, SOC exhibits opposite filling-dependent behaviors for the
superfluidity\cite{arxivsun2013spin}. These interesting physics induced by SOC
are all investigated by the Bogoliubov-de Gennes (BdG) approach. Moreover, the
study of the spin-orbit coupled Fermi gases in lattice at finite temperature is
still waiting to be explored.

Two effects are resulted by applying SOC in the Fermi Hubbard model. First, SOC
enhances the effective hopping amplitude and enlarges the bandwidth. The other
is that SOC flips the spin of the fermion which breaks the rotational symmetry
of the spin and significantly changes the properties of the Fermi surface. When
the system only contains the SOC, the ground state is semimetal near
half-filling\cite{arxivsun2013spin} with vanishingly small density of
state(DOS)($\rho(E) \sim |E|$). In the strong attractive limit, the fermions
are strongly bounded and the superfluid transition temperature is determined by
the center-of-mass motion which is proportional to the inverse of the
attraction. Therefore, our major concern here is to investigate what effects
can be induced by the  SOC on the pairing at finite temperature beyond the BdG
approach.

In this Letter we investigate the pairing of the attractive Fermi gases in 2D
square optical lattice with SOC using both DQMC
simulations\cite{PhysRevB.38.12023,PhysRevB.37.5070,PhysRevLett.62.1407,PhysRevLett.104.066406,PhysRevLett.69.2001}
and mean field theory. To our knowledge, this is the first unbiased numeric
simulation of the  spin-orbit coupled Fermi gases. Our results give us a
detailed description about the pairing behavior and the superfluid phase
transition of this  spin-orbit coupled system at finite temperature. The main
results are summarized as following: (1) With  SOC, there exists
Berezinskii-Kosterlitz-Thoules (BKT) phase transition even in the absence of
the hopping term. The superfluid phase transition temperature is enhanced by
SOC in strong attraction region. In intermediate region, the superfluid
transition temperature is always suppressed by SOC. The peak of the transition
temperature is approximately proportional to the bandwidth, which is enlarged
significantly by large SOC. (2) SOC always suppresses the pairing temperature
at strong attraction region. Thus, SOC has an opposite effect on the pairing
and the superfluidity in this region. These are qualitatively different from
the continuous case\cite{PhysRevLett.108.025301,PhysRevLett.109.105302}. (3)
Due to the emergence of spin-triplet pairing and the breaking of the rotational
symmetry of spin, the spin susceptibility becomes anisotropic. When the
temperature decreases to zero, spin susceptibility retains finite values.

\textit{Model and Method:} We start with the 2D Rashba spin-orbit coupled
fermionic Hubbard model on a square lattice which can be written as following.
\begin{eqnarray}
H&=&-t\sum_{\langle{i,j}\rangle}c^{\dagger}_{i,s}c_{j,s}+ i\lambda\sum_{\langle{i,j}\rangle} c_{i,s}^{\dagger}(\mathbf{e}_{i,j} \times \boldsymbol{\sigma})^{s,s'}_{z}c_{j,s'}\nonumber\\
&&-U\sum_{i}n_{i,\uparrow}n_{i,\downarrow}-\mu\sum_{i}n_{i},
\label{hamiltonian}
\end{eqnarray}
where $c^{\dagger}_{i,s}(c_{i,s})$ denotes the creation (annihilation)
operators for fermionic atoms with spin $s \equiv (\uparrow,\downarrow)$ at
site $i$. $n_{i}$ is the fermionic density operator at site $i$:
$n_{i}=\Sigma_{s}n_{i,s}=\Sigma_{s}c^{\dagger}_{i,s}c_{i,s}$.
$\boldsymbol{\sigma}$ is the Pauli matrices, $\mathbf{\hat{e}}_{i,j}$ is the
vector connecting sites $i$ and $j$.  $\langle{i,j}\rangle$ denotes the
summation over the nearest neighbors. $t$, $\lambda$, $U(U>0)$ and $\mu$ stand
for the hopping amplitude, Rashba SOC strength, on-site attractive interaction,
and chemical potential, respectively.

SOC lifts the spin degeneracy and gives rise to two splited helical branches
for noninteracting case. The two helical branches have four contact Dirac cones
at $[(0,0),(0,\pi),(\pi,0)$ and $(\pi,\pi)]$. The splitting between two
branches increases with SOC. The bandwidth is enlarged and the half bandwidth
is $W(t,\lambda)=4t \sqrt{\frac{2}{(2  + \lambda^2/t^{2})}} +2 \lambda
\sqrt{\frac{2 \lambda^2/t^{2}}{2 + \lambda^2/t^{2}}}$. The DOS diverges at four
van Hove singularities $\omega=\pm 2t  \pm 2 t \sqrt{1 + \lambda^{2}/t^{2}}$
instead at $\omega = 0$ \cite{Supplementary}.  At half filling, the Hamiltonian
has a particle-hole symmetry with $c_{i,s} \rightarrow
d_{i,s}^{\dag}=(-1)^{i_{x}+i_{y}} c_{i,s}$ and $c_{i,s}^{\dag} \rightarrow
d_{i,s}=(-1)^{i_{x}+i_{y}} c_{i,s}^{\dag}$. The Fermi surface is perfectly
nested with a nesting vector $Q=(\pi,\pi)$. Throughout this paper, we use the
hopping amplitude $t$ as the unit energy and assume $t=1$.

Since the SOC is a complex spin-flip term, the BSS algorithm of DQMC should be
modified to updates the up-spin and down-spin simultaneously instead of
updating them separately. By this modification, the notorious sign-problem
becomes more troublesome. Fortunately, our model is free of sign-problem in the
DQMC simulations. This guarantees our DQMC simulations to achieve a good
numerical precision at large size and low temperature. Typical system in our
DQMC simulations is $10 \times 10$ and periodic boundary condition, the
Suzuki-Trotter decomposition (the step is $\Delta \tau = \beta / M = 0.125$
with $\beta = 1/T$) is used and then a discrete Hubbard-Stratonovich
transformation is introduced to decouple the on-site attractive interaction
into a bilinear form. The systematic error of our DQMC simulations on the order
of $(\Delta \tau)^2$.

\textit{Berezinskii-Kosterlitz-Thoules phase transition:} In two dimensions,
although pairs can be formed, there is no long-range superfluid order at finite
temperature because of the spatially-dependent phase fluctuation, so there is
no condensation. At finite temperature, BKT phase transition is possible for
the emergence of quasi-long-range (algebraic long-range) superfluid order. When
temperature drops below a critical temperature ($T_{BKT}$), the system
undergoes a phase transition from the pseudogap phase to the superfluid phase.
On the two sides of $T_{BKT}$, the superfluid density has a universal jump. The
$T_{BKT}$ can be precisely determined by this
jump\cite{PhysRevLett.39.1201,Supplementary}.
\begin{eqnarray}
T_{BKT}=\frac{\pi}{2}D_{s}(\lambda,U,T_{BKT})
\end{eqnarray}
where $D_{s}(\lambda,U,T_{BKT})$, which can be determined by the
current-current correlation function\cite{PhysRevLett.68.2830}, is the
superfluid density  at the superfluid side of $T_{BKT}$.

\begin{figure}
\includegraphics[width=\linewidth]{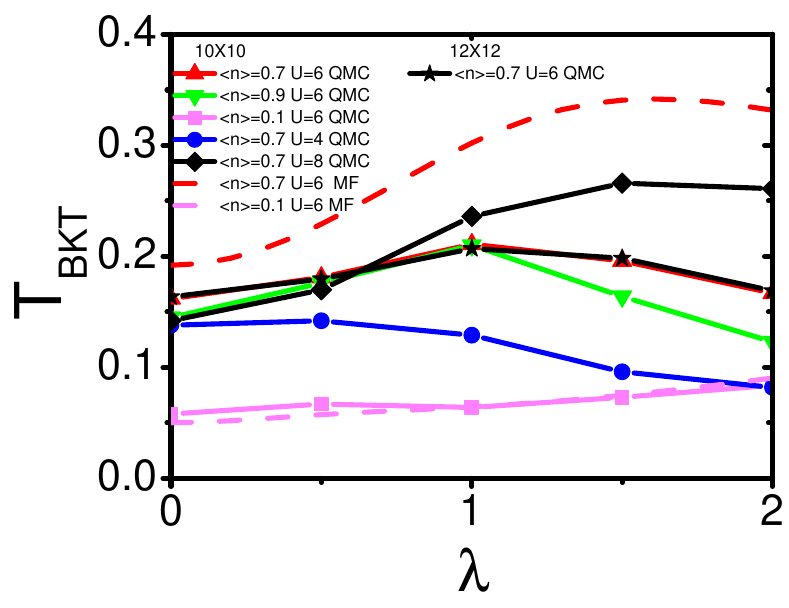}
\caption{$T_{BKT}$ VS $\lambda$ for various $\langle{n}\rangle$ and $U$. The
solid curves represent the results of DQMC, while the dashed curves is the
results of mean field theory. $T_{BKT}$ is enhanced firstly and then suppressed
by SOC at strong attraction($U=6,8$). Whereas $T_{BKT}$ is always suppressed by
SOC at intermediate attraction($U=4$). The $T_{BKT}$ curves at the size
$12\times12$ and $10\times 10$ almost coincide for $\langle n\rangle=0.7$ with
$U=6$ case. The results of mean field and DQMC are consistent quantitatively
for small filling and only qualitatively for large filling.} \label{Fig1}
\end{figure}

In Fig.\ref{Fig1}, we show $T_{BKT}$ as a function of $\lambda$ for different
$\langle{n}\rangle$ with $U=4,6,8$. We also have performed the DQMC simulations
on $12\times12$ lattice size for $U=6$ with $\langle n\rangle=0.7$ case.
$T_{BKT}$ curve of $12 \times 12$ lattice size almost coincides with the curve
of $10 \times 10$ size as shown in Fig.\ref{Fig1}. Thus, our simulations are
credible for $10 \times 10$ lattice size. For $U=6,8$ cases, $T_{BKT}$ is
firstly enhanced and then suppressed by SOC, whereas $T_{BKT}$ is always
suppressed by SOC for $U=4$ case. These are resulted by the competition between
the pair breaking and the center-of-mass motion. In strong attraction
case($U>z$, $z$ is coordinate number), the fermions form tight cooper pairs and
$T_{BKT}$ is controlled by the center-of-mass motion.  When SOC increases, the
center-of-mass motion is enhanced due to the enlargement of the bandwidth.
Therefore, $T_{BKT}$ will be enhanced.  When SOC becomes larger than a critical
value $\lambda_{c}$, the pair breaking would be dominant comparing to the
enhancement of the center-of-mass motion, then SOC would suppress $T_{BKT}$.
When $U$ increases, the cooper pairs will become tighter, and thus
$\lambda_{c}$ will increases. For $U=4$ case, this is an intermediate region
between the strong and weak attraction region. The cooper pairs are more
loosely formed. Therefore, the $T_{BKT}$ will be suppressed by increasing SOC.
For weak attraction case($U<z$), $T_{BKT}$ is too low to be exactly determined
by DQMC simulations. In mean field framework, we find that $T_{BKT}$ always
decreases with increasing SOC for large filling case($0.5<\langle n\rangle <1$)
and non-monotonous decreases for small filling case. This behavior is dominated
by the Fermi surface density of state\cite{Supplementary}. Fig.\ref{Fig1} also
shows that $T_{BKT}$ increases with filling. However $T_{BKT}$ will drop
rapidly when  $\langle n\rangle$ approach to 1 due to the stability of
charge-density wave. This is resemble to the case without
SOC\cite{PhysRevLett.62.1407}. We also show the mean field results in
Fig.\ref{Fig1}. We find that the results of the two methods are consistent
quantitatively at small filling ($\langle{n}\rangle=0.1$) and qualitatively at
large filling($\langle{n}\rangle=0.7$) for $U=6$.

\begin{figure}
\includegraphics[width=\linewidth]{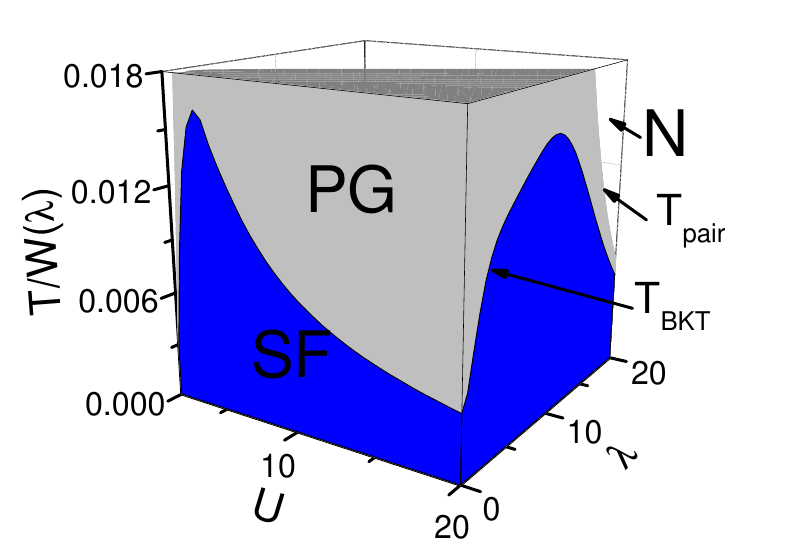}
\caption{The finite temperature phase diagram on the $U-\lambda$ plane
determined by mean field for $\langle{n}\rangle=0.1$. The temperature is in the
unit of half bandwidth $W(\lambda)$ which depends on $\lambda$. There are three
regions in the phase diagram- superfluid(SF), pseudogap(PG) and normal(N). The
boundary between PG and N is estimated by the vanishing of pairing amplitude.
The peak of $T_{BKT}$ is approximately proportional to $W(\lambda)$.}
\label{Fig2}
\end{figure}

To visualize the effect of SOC and $U$ on $T_{BKT}$ and $T_{pair}$, we give the
finite temperature phase diagram in Fig.\ref{Fig2} for $\langle{n}\rangle=0.1$
obtained from the mean field theory. The half bandwidth $W(\lambda)$ is used as
the unit of temperature. There are three regions in the phase diagram. From
high temperature to zero temperature, the phases are normal(N), pseudogap(PG)
and superfluid(SF) phase. The maximum $T_{BKT}$  is approximately proportional
to the bandwidth $T_{BKT}^{max}\simeq c(\langle{n}\rangle) W(\lambda)$. Thus,
$T_{BKT}$ can be significantly enhanced by large SOC. This also indicates that
there exists finite temperature superfluid phase transition even in the absence
of the hopping term. The PG region is determined by nonzero pairing amplitude
without superfluidity. Fig.\ref{Fig2} intuitively reveals behavior of the
$T_{BKT}$ and $T_{pair}$. $T_{pair}$ is always suppressed by SOC at strong
attraction region. This is qualitatively different from the continuous case
\cite{PhysRevLett.108.025301,PhysRevLett.109.105302}. In continuous case, SOC
always suppresses the superfluidity but enhances the pairing.

\textit{Pairing susceptibilities:} As we mentioned above, SOC breaks the spin
rotational symmetry, so the pairing symmetry will also be changed. To
investigate the symmetry of the pairing, we calculated the zero frequency
($q=0$, $\omega=0$) pair susceptibilities.

\begin{eqnarray}
P_{\gamma}=\int_{0}^{\beta} d \tau \langle\Delta_{\gamma}(\tau) \Delta^{\dagger}_{\gamma}(0)\rangle,
\end{eqnarray}
where $\gamma$ denotes the pairing symmetry. The spin-singlet pairing and spin-triplet pairing are
\begin{eqnarray}
&&\Delta_{s\uparrow\downarrow} =\sum_k c_{k,\downarrow} c_{-k,\uparrow},\\
&&\Delta_{p\uparrow\uparrow} =\sum_k (\sin k_{x}+ \sin k_{y})c_{k,\uparrow} c_{-k,\uparrow}.
\end{eqnarray}

The pair susceptibilities diverge at a critical temperature below which the
system has a quasi-long range superfluid order and undergo a superfluid phase
transition. By introducing the uncorrelated pair susceptibilities
$\widetilde{P}_{\gamma}$, the interaction vertex is
$\Gamma_{\gamma}=\frac{1}{P_{\gamma}}-\frac{1}{\widetilde{P}_{\gamma}}$
\cite{PhysRevB.39.839}. The pairing channel is attractive for negative pairing
vertex ($\Gamma_{\gamma} \cdot \widetilde{P}_{\gamma}<0$), while it is
repulsive for positive pairing vertex ($\Gamma_{\gamma} \cdot
\widetilde{P}_{\gamma}>0$). Superfluid instability is signaled by
$\Gamma_{\gamma} \cdot \widetilde{P}_{\gamma}\rightarrow -1$.

\begin{figure}
\includegraphics[width=\linewidth]{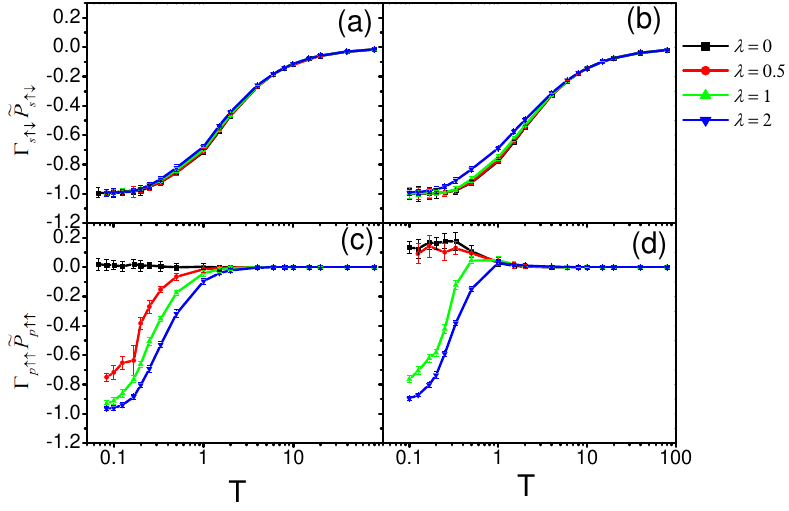}
\caption{The pairing vertex as a function of temperature for spin-singlet
pairing: (a)$\langle{n}\rangle=0.1$, (b)$\langle{n}\rangle=0.7$ and
spin-triplet pairing: (c)$\langle{n}\rangle=0.1$, (d)$\langle{n}\rangle=0.7$
with $U=6$ and different $\lambda$. The convergence is decelerated by  SOC for
spin-singlet pairing while is accelerated for spin-triplet pairing}
\label{Fig3}
\end{figure}

Fig.\ref{Fig3} shows the pairing vertex of spin-singlet(up row) and
spin-triplet(down row) with $\lambda=0,0.5,1,2$, and $U=6$ versus temperature.
Fig.\ref{Fig3}(a) and (b) show that spin-singlet pairing vertex converge to
$-1$ as $T \rightarrow 0$ and contribute to superfluid. While the triplet
pairing vertex converge to $0$ in the absence of  SOC and converge to $-1$ in
the presence of  SOC (except Fig.\ref{Fig3}(d) $\lambda=0.5$ case). This
indicates that spin-triplet pairing can emerges and contributes to superfluid
in the presence of  SOC. The pairing of the superfluid is a mixture of
spin-singlet and spin-triplet. They compete with each other in the system as
the  SOC increases. In Fig.\ref{Fig3}(d), pairing vertex does not converge to
$-1$ for $\lambda=0.5$. This indicates that there exists a critical  SOC
strength($\lambda_{c}$) above which spin-triplet pairing has contribution to
superfluid. The convergence is decelerated by  SOC for spin-singlet pairing
while is accelerated for spin-triplet pairing.

\textit{Spin susceptibilities:} Because the symmetry of the pairing has been
changed by  SOC, the spin response will be very different from the case without
SOC, especially the spin susceptibility. Without the  SOC, the pairing is only
spin-singlet and the spin susceptibility is isotropic. When the temperature
decreases, thermodynamic fluctuation will be suppressed and this will render
the enhancement of spin susceptibility. When the temperature decreases to a
critical value, spin-singlet pairs are formed, so the spin susceptibility will
be suppressed. When the temperature decreases to zero, all the fermions are
paired. Thus, spin susceptibility would decrease to zero
\cite{PhysRevLett.104.066406,PhysRevLett.69.2001}. In the presence of  SOC, the
spin susceptibility becomes anisotropic and can be written as following:
\begin{equation}
\chi_{\alpha} = \frac{1}{N} \sum_{i,j} e^{-i \vec{q} \cdot
(\vec{r}_i-\vec{r}_j)}\int_0^{\beta} d \tau <s_{i}^{\alpha}(\tau)\cdot
s_{j}^{\alpha}(0)> |_{\vec{q} \rightarrow 0},
\end{equation}
where $s^{\alpha}$ is the spin with $\alpha=(x,y,z)$.

\begin{figure}
\includegraphics[width=\linewidth]{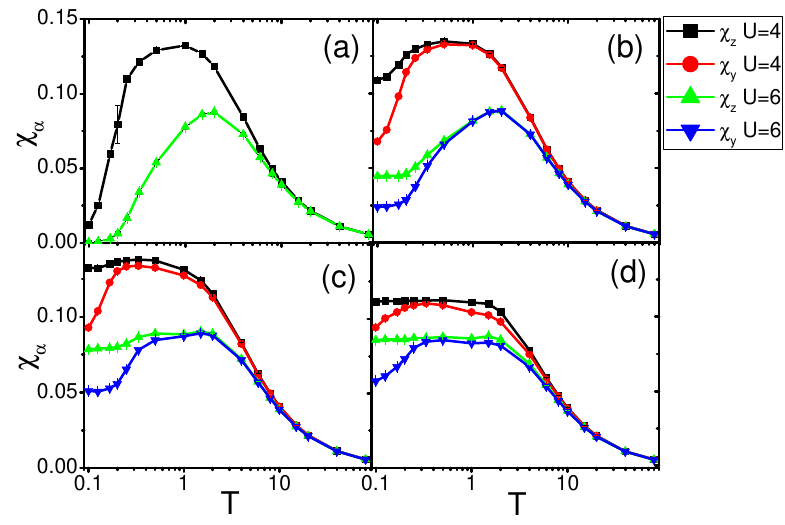}
\caption{The spin susceptibilities vs temperature with
$\lambda=$(a)0,(b)0.5,(c)1,(d)2 and $\langle{n}\rangle=0.7$. When the
temperature decreases to zero, spin susceptibilities tend to finite values for
$\lambda \neq 0$ and 0 for $\lambda = 0$.} \label{Fig4}
\end{figure}

Fig.\ref{Fig4} shows spin susceptibilities as functions of temperature with
$\lambda=0,0.5,1,2$, $U=4,6$ and $\langle{n}\rangle=0.7$.  The curves of spin
susceptibilities are smooth. The spin susceptibilities remain unchanged across
$T_{BKT}$. For $\lambda=0$ our result agrees with the
Ref\cite{PhysRevLett.104.066406} as shown in Fig.\ref{Fig4}(a). In the presence
of SOC, the anisotropic spin susceptibilities as shown in Fig.\ref{Fig4}(b)-(d)
for different $\lambda$. When the temperature decreases, spin susceptibilities
increase firstly and then gradually decrease. Significantly different from the
$\lambda=0$ case, the spin susceptibilities does not drop to zero but remains
finite even when temperature decreases to zero. This can be understood by the
formation of spin-triplet pairing. Spin-singlet pairing has zero total spin and
has no contribution to the spin susceptibilities unless being broken by
thermodynamic fluctuation. Quite the contrary, the spin-triplet pairing
possesses total spin and contributes to spin susceptibilities even at zero
temperature. Thus, the spin susceptibilities retain finite values when
temperature approaches to zero. The finite values of spin susceptibilities
reveal the weight of spin-triplet pairing. The spin susceptibilities are also
suppressed by attraction $U$ for the on-site attraction favors the spin-singlet
pairing. Certainly, we can also estimate the pairing temperature $T_{pair}$
from Fig.\ref{Fig4} by the location of the peak of $\chi$. $T_{pair}$ is
approximately equal to $1$ which is much larger than $T_{BKT}$. Therefore,
there is a large pseudogap region in finite temperature phase diagram which
confirms the validity of the mean field phase diagram in Fig.\ref{Fig2}.  As
for the spin-triplet pairing, here $T_{pair}$ is underestimated.

\begin{figure}
\includegraphics[width=\linewidth]{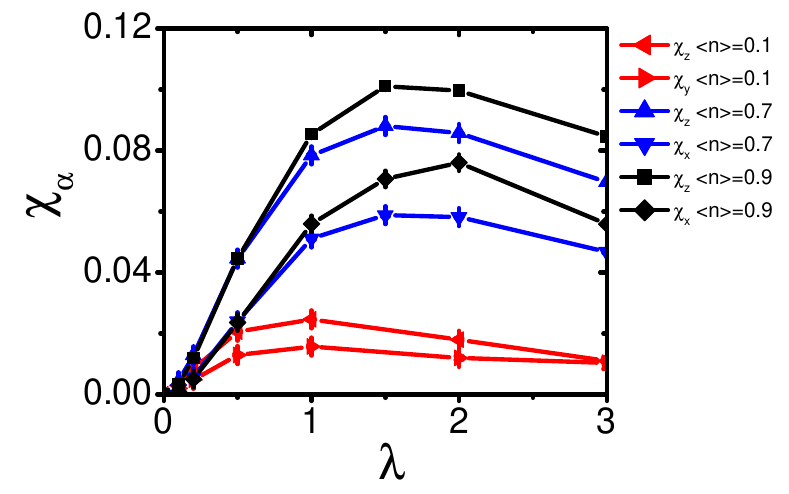}
\caption{Spin susceptibilities $\chi_{\alpha}$ vs $\lambda$ with $U=6$ and
$\langle{n}\rangle=0.1,0.7,0.9$ for $\beta=1/T=10$. At small  SOC limit, spin
susceptibilities are the quadratic functions of $\lambda$.} \label{Fig5}
\end{figure}

Fig.\ref{Fig5} shows the spin susceptibilities as functions of $\lambda$ for
different $\langle{n}\rangle$ at $\beta=10$ with $U=6$. Spin susceptibilities
increase firstly with $\lambda$ for the increasing of the spin-triplet pairing.
In small SOC limit, spin susceptibilities are the quadratic functions of
$\lambda$ which is in accord with the continuous
case\cite{PhysRevLett.87.037004}. At large  SOC, the spin susceptibilities are
suppressed for the reason that SOC suppresses the pairing of both spin-singlet
and spin-triplet as discussed above.

\textit{Discussion and Conclusion:} Obviously, our DQMC simulations and the
results could be applicable to the La AlO$_{3}$/SrTiO$_{3}$
interface\cite{PhysRevLett.107.056802,bertnaturephysics2011direct,banerjeenaturephy2013ferromagnetic}
and noncentrosymmetric superconductors such as CePt$_{3}$Si,
Li$_{2}$(Pt$_{1-x}$Pd$_{x}$)$_{3}$B\cite{PhysRevLett.97.017006,PhysRevB.87.161203},
because strong SOC exists in these materials. The behavior of spin
susceptibilities can be determined by Knight shift in nuclear magnetic
resonance(NMR) measurements\cite{PhysRevLett.98.047002}.

We have performed simulations for the attractive fermionic Hubbard model with
Rashba SOC in 2D square optical lattice using DQMC and mean field theory. There
exists a finite temperature superfluid phase transition. The transition
temperature is suppressed by SOC in intermediate attraction. With the strong
attraction, the superfluid transition temperature is enhanced firstly and then
suppressed by SOC. The spin susceptibility becomes anisotropic and retains
finite values when the temperature approach to zero. This nontrivial behavior
of spin susceptibilities can be confirmed by speckle
imaging\cite{PhysRevLett.106.010402} in experiments. We also check the
anisotropic SOC case which can be consider as a mixture of Rashba and
Dreselhaus SOC. We find that the behavior of superfluid transition temperature
resemble to the Rashba SOC case while the isotropic of spin susceptibility in
$x-y$ plane will be further destroyed.

\textit{Acknowledges:} We would like to thank Prof. W. Yi, Youjin Deng, Hui
Zhai, Tianxing Ma and Q. Sun for helpful discussions. H.-K. Tang would like to
thank the supports from Prof. ShiJian Gu. This work is supported by NSFC
91230203, CAEP, and China Postdoctoral Science Foundation (No. 2012M520147).

\bibliography{reference}

\begin{widetext}
\section{Supplementary Material}
In this supplementary material, we present some details of the calculations.

\subsection{van Hove singularity}
The single-particle Hamiltonian
$H_{0}=-t\sum_{\langle{i,j}\rangle}c^{\dagger}_{i,s}c_{j,s}+
i\lambda\sum_{\langle{i,j}\rangle} c_{i,s}^{\dagger}(\mathbf{e}_{i,j} \times
\boldsymbol{\sigma})^{s,s'}_{z}c_{j,s'}$. The dispersion of the two helical
branches are $\epsilon_{\mathbf{k},\nu=\pm}=-2t(\cos k_{x} + \cos k_{y})+  2
\lambda \nu \sqrt{\sin ^2 k_{x} +\sin ^2 k_{y}}$. The van Hove singularity is
\begin{eqnarray}
  &&|\nabla \epsilon _{\mathbf{k},\nu}| = 0\nonumber\\
                           &&\Rightarrow\left\{\begin{array}{c}
                                     \sin k^{c}_{x}=0~~ or~~ \cos k^{c}_{x} = - \nu t\sqrt{\sin ^{2} k^{c}_{x} + \sin^{2} k^{c}_{y}}/\lambda \\
                                     \sin k^{c}_{y}=0 ~~or~~ \cos k^{c}_{y} = - \nu t\sqrt{\sin ^{2} k^{c}_{x} + \sin^{2} k^{c}_{y}}/\lambda
                                   \end{array}
                            \right.\nonumber
\end{eqnarray}

There are three types of van Hove singularities: (I) $\sin k^{c}_{x}= \sin
k^{c}_{y}=0$ with $\epsilon_{\mathbf{k}^{c}}=\pm4t,0$; (II) $\sin k^{c}_{x}=0,
\cos k^{c}_{y} = - \nu t \sqrt{\sin ^{2} k^{c}_{x} + \sin^{2}
k^{c}_{y}}/\lambda$ (or $\sin k^{c}_{y}=0, \cos k^{c}_{x} = - \nu t \sqrt{\sin
^{2} k^{c}_{x} + \sin^{2} k^{c}_{y}}/\lambda$) with
$\epsilon_{\mathbf{k}^{c}}=\pm2t \pm 2t\sqrt{1+(\lambda/t)^{2}}$; (III) $\cos
k^{c}_{x}=\cos k^{c}_{y} =- \nu t \sqrt{\sin ^{2} k^{c}_{x} + \sin^{2}
k^{c}_{y}}/\lambda$ with $\epsilon_{\mathbf{k}^{c}}=\pm 4t
\sqrt{\frac{2}{2+(\lambda/t)^{2}}} \pm 2t
\sqrt{\frac{2(\lambda/t)^{2}}{2+(\lambda/t)^{2}}}$.

The half bandwidth is $W(t,\lambda)= 4t \sqrt{\frac{2}{2+(\lambda/t)^{2}}} + 2
\lambda \sqrt{\frac{2(\lambda/t)^{2}}{2+(\lambda/t)^{2}}}$ which increases with
SOC.

The divergence of DOS only comes from the narrow region which contains the
$\mathbf{k}^{c}$. Thus, we dive the integral into two parts: $a$ labels the
narrow region which contains $\mathbf{k}^{c}$ and $b$ labels the other region
of the integral. The DOS at van Hove singularities is
\begin{eqnarray}
  \rho (\epsilon_{\mathbf{k}}^{c}) &=& \frac{1}{N}\sum_{\mathbf{\mathbf{k}},\nu} \delta[\epsilon_{\mathbf{\mathbf{k}}^{c}}- \epsilon_{\mathbf{k},\nu}]\nonumber\\
  &=& \sum_{\nu} \int_{0}^{\pi}\frac{d k_{x} d k_{y}} {\pi^2}\delta[\epsilon_{\mathbf{k}^{c}}-\epsilon_{\mathbf{k},\nu}]\nonumber\\
  &=& \sum_{\nu} \int_{a+b}\frac{d k_{x} d k_{y}} {\pi^2}\delta[\epsilon_{\mathbf{\mathbf{k}}^{c}}-\epsilon_{\mathbf{\mathbf{k}},\nu}]\nonumber
\end{eqnarray}
We only consider the integral in $a$ region that contributes the divergence of
the DOS.  At this narrow region, the dispersion can be expanded as
$\epsilon_{\mathbf{k},\nu}=\epsilon_{\mathbf{k}^{c}}
+\epsilon_{\mathbf{k}',\nu}$  with $\mathbf{k}=\mathbf{k}^{c}+ \mathbf{k}'$
($k'_{x,y} = [0,\Lambda], \Lambda << \pi$).
\begin{eqnarray}
  \rho (\epsilon_{\mathbf{k}^{c}}) = \sum_{\nu} \int_{0}^{\Lambda}\frac{d k'_{x} d k'_{y}} {\pi^2} \delta[\epsilon_{\mathbf{k}',\nu}]\nonumber
\end{eqnarray}
with
\begin{eqnarray}
  \epsilon_{\mathbf{k}',\nu}=\left\{\begin{array}{c}
                                       c_{1} \sqrt{k_{x}^{'2} + k_{y}^{'2}}~~~~\text{for~ I~case}\\
                                       c_{2,x}k_{x}^{'2} - c_{2,y} k_{y}^{'2}~~ \text{for~ II~case}\\
                                       c_{3}k_{x}^{'2} + c_{3} k_{y}^{'2}~~~~~~~ \text{for~ III~case}
                                    \end{array}
   \right.  \nonumber
\end{eqnarray}
where, $c_{1}=2 \nu\lambda$, $c_{2,x}=(t \cos k^{c}_{x} + \nu
t\sqrt{1+(\lambda/t)^{2}})$, $c_{2,y}=\nu (\lambda^{2}/t +
t)/\sqrt{1+(\lambda/t)^2}$, $c_{3}= -\nu \frac{\sqrt{2}
(t^2+\lambda^2)}{\sqrt{2 t^2+\lambda^2}}$. Here
$\text{sgn}(c_{2,x})=\text{sgn}(c_{2,y})$. Therefore, DOS logarithmical
diverges for II case and converges for I and III cases.

At the bottom(III case) of the dispersion, the effective mass is
$\frac{1}{[m^{*}]_{ij}}=\frac{\partial \epsilon_{\mathbf{k},\nu}}{\partial
k_{i} \partial k_{j}}|_{\mathbf{k}^{c}}$ with $i,j=(x,y)$.

\begin{eqnarray}
\frac{1}{m^{*}_{xx}}=\frac{1}{m^{*}_{yy}}&=&t\cos k^{c}_{x}+\sqrt{2}\lambda \sin k^{c}_{x}\nonumber\\
\frac{1}{m^{*}_{xy}}=\frac{1}{m^{*}_{yx}}&=&t\cos k^{c}_{x}\nonumber
\end{eqnarray}

Then, $\frac{2}{m^{*}_{xx}}+ \frac{2}{m^{*}_{xy}} = W(t,\lambda)$. Therefore,
the effective mass is suppressed by increasing SOC.

\begin{figure}
\includegraphics[width=\linewidth]{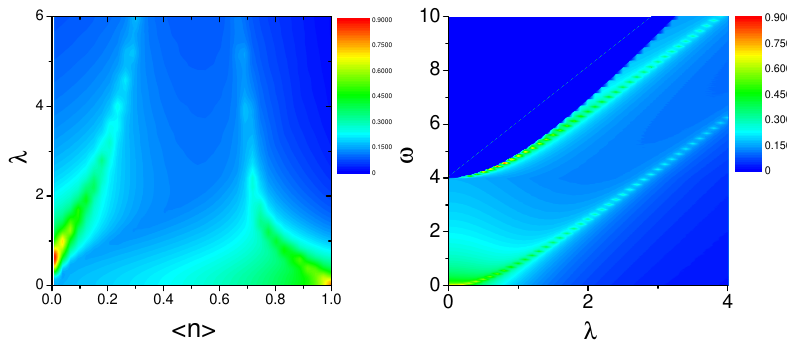}
\caption{Left is the DOS at Fermi surface on $\langle n\rangle$-$\lambda$
plane. Right is the DOS on $\lambda$-$\omega$ plane.} \label{Figdos}
\end{figure}

If the system only contains the SOC term, the two helical branches dispersion
of the single-particle are $\epsilon_{\mathbf{k},\nu=\pm}= 2 \lambda \nu
\sqrt{\sin ^2 k_{x} +\sin ^2 k_{y}}$. Near half filling, the dispersion can be
expanded at $(0,\pi)$ and $(\pi,0)$ points. Then the Fermi surface DOS is $\rho
(E)=\sum_{k,\nu}\delta (E-2 \nu \lambda \sqrt{k_{x}^{2}+k_{y}^{2}})\sim
|E|/\lambda$.

\subsection{BKT transition temperature}
At finite temperature, the spatially-dependent phase fluctuation will always
breaks the long-rang order in two dimensions. The vortex like phase fluctuation
can induce a phase transition between the algebraic long-rang order
(quasi-long-rang order) and the short-rang order. This is the BKT phase
transition. The superfluid density has a universal jump and can be determined
by current-current correlation. Here, we give the derivation of current formula
by linear response. The current formula can also
be directly derived by $\vec{J}=i[H,\vec{P}]$ with the polarization operator
$\vec{P}=\sum_{i}\vec{R}_{i} n_{i}$. In the presence of a small vector
potential $A_{x}(i)$, the hopping and the SOC term are modified by a Peierls
phase
\begin{eqnarray}
H^{A}_{0}&=&-t \sum_{i,s} \left[c_{i+x,s}^{\dag}c_{i,s}e^{ie A_{x}(i)} + c_{i,s}^{\dag}c_{i+x,s}e^{-i e A_{x}(i)} +c_{i+y,s}^{\dag} c_{i,s} + c_{i,s}^{\dag} c_{i+y,s}\right] \nonumber\\
&&- \lambda \sum_{i}\left[(c_{i-x,\downarrow}^{\dag} c_{i,\uparrow} e^{-ie A_{x}(i)}- c_{i+x,\downarrow}^{\dag} c_{i,\uparrow} e^{ie A_{x}(i)}) + i (c_{i-y,\downarrow}^{\dag} c_{i,\uparrow}- c_{i+y,\downarrow}^{\dag} c_{i,\uparrow}) + H.c.\right].
\end{eqnarray}
The Hamiltonian can be expanded in the order of the small vector potential.
\begin{eqnarray}
H^{A}_{0}= H_{0} + \bar{H}_{0}^{A},
\end{eqnarray}

where
\begin{eqnarray}
\bar{H}_{0}^{A}=-\sum_{i}\left[e J_{x}^{P} A_{x}(i) + \frac{e^2 A_{x}^{2}(i)}{2} K_{x}(i)\right]\nonumber
\end{eqnarray}
with
\begin{eqnarray}
J_{x}^{P}&=&i t \sum_{i,s}(c_{i+x,s}^{\dag}c_{i,s}-c_{i-x,s}^{\dag}c_{i,s})+i \lambda \sum_{i}(c_{i-x,\downarrow}^{\dag} c_{i,\uparrow} + c_{i+x,\downarrow}^{\dag} c_{i,\uparrow})- i \lambda \sum_{i} (c_{i,\uparrow}^{\dag} c_{i-x,\downarrow}+ c_{i,\uparrow}^{\dag} c_{i+x,\downarrow})\nonumber\\
K_{x}(i)&=&- t\sum_{i,s} (c_{i+x,s}^{\dag} c_{i,s} + c_{i,s}^{\dag}c_{i+x,s}) - \lambda \sum_{i} (c_{i-x,\downarrow}^{\dag}c_{i,\uparrow}-c_{i+x,\downarrow}^{\dag}c_{i,\uparrow}) -\lambda \sum_{i} (c_{i,\uparrow}^{\dag}c_{i-x,\downarrow}-c_{i,\uparrow}^{\dag}c_{i+x,\downarrow})\nonumber
\end{eqnarray}

The current-current correlation is
\begin{eqnarray}
\Lambda_{xx} (q,i \omega_{m})= \int_{0}^{\beta} d \tau e^{i \omega_{m} \tau}\langle J_{x}^{P}(q,\tau) J_{x}^{P}(-q,0)\rangle.
\end{eqnarray}

Then, the superfluid density is:
\begin{eqnarray}
D_{s}(T) = \frac{1}{4}\Big[<-K_x> - \Lambda_{xx} (q_x=0,q_y \rightarrow 0,i \omega_m =0)\Big].
\end{eqnarray}

The BKT transition temperature satisfies
\begin{eqnarray}
T_{BKT} = \frac{\pi}{2}D_{s}(T_{BKT}).
\end{eqnarray}

\begin{figure}
\includegraphics[width=\linewidth]{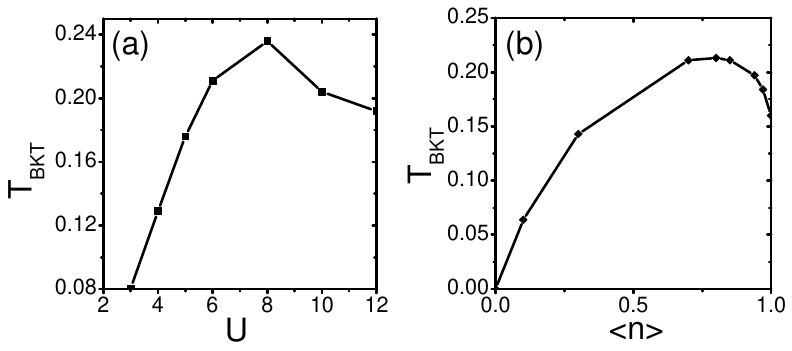}
\caption{$T_{BKT}$ as a function of (a)$U$ with $\lambda=1$ and $\langle
n\rangle =0.7$, (b)$\langle n\rangle$ with $U=6$ and $\lambda=1$. $T_{BKT}$
drop rapidly as $\langle n\rangle$ approach to half filling.} \label{tbkt}
\end{figure}

\subsection{Mean field framework}
In mean field framework, the partition function of our system can be written as
following by introducing the basis
$\psi_{i}=(c_{i,\uparrow},c_{i,\downarrow},c^{\dag}_{i,\uparrow},c^{\dag}_{i,\downarrow})^{T}$.
\begin{eqnarray}
\textit{Z}=\int \textit{D}[\bar{\psi},\psi] e^{-S[\bar{\psi},\psi]},
\end{eqnarray}
where the action is
\begin{eqnarray}
S[\bar{\psi},\psi]=\int_{0}^{\beta} d \tau \left[\sum_{s} \bar{\psi} \partial_{\tau} \psi + H(\bar{\psi},\psi)\right].
\end{eqnarray}

With the Hubbard-Stratonovich transformation $\Delta_{i}=-U\langle
c_{i,\downarrow} c_{i,\uparrow}\rangle$ and integrating out the fermion degrees
of the freedom, we have the partition function $\textit{Z}=\int
\textit{D}[\bar{\Delta},\Delta] e^{-S_{eff}[\bar{\Delta},\Delta]}$ with the
effective action
\begin{eqnarray}
S_{eff}[\bar{\Delta},\Delta]=\int_{0}^{\beta} d \tau \Big( |\Delta|^{2}/U + \varepsilon_{k}\Big) - \frac{1}{2} \text{Tr}[\ln G^{-1}].
\end{eqnarray}
Here, the inverse Green function is
\begin{eqnarray}
G^{-1}=\left(
         \begin{array}{cc}
           \partial_{\tau}+\varepsilon_{k} + g_{k}  & - i \Delta \sigma_{y} \\
           i \bar{\Delta} \sigma_{y} & \partial_{\tau}- \varepsilon_{k} + g_{k}^{T} \\
         \end{array}
       \right),\label{greenfunction}
\end{eqnarray}
with $\varepsilon_{k} = \epsilon_{k} - \mu$, $\epsilon_{k}=- 2 t (\cos k_{x}+
\cos k_{y})$ and $g_{k} = 2 \lambda (\sin k_{y} \sigma_{x} - \sin k_{x}
\sigma_{y})$.

If we ignore the spatial-dependent phase fluctuation $\Delta_{i} = \Delta$, we
have the gap and the number equations as following:
\begin{eqnarray}
\frac{1}{U} &=& \sum_{k,\nu=\pm}\frac{1}{4E_{k,\nu}}\left[1-2f(E_{k,\nu})\right],\nonumber\\
n &=& \frac{1}{2}\sum_{k,\nu=\pm}\left[1-\frac{\varepsilon_{k,\nu}}{E_{k,\nu}}(1-2f(E_{k,\nu}))\right].\nonumber
\end{eqnarray}
Here, the excitation spectrum is $E_{k,\nu}=\sqrt{\varepsilon_{k,\nu}+\Delta
^{2}}$ with $\varepsilon_{k,\nu}=\varepsilon_{k} +\nu |g_{k}| $ . The pairing
temperature is determined by $\Delta=0$.

\begin{figure}
\includegraphics[width=\linewidth]{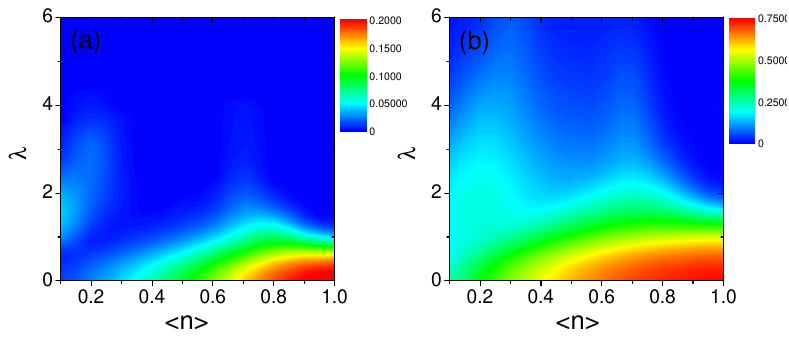}
\caption{$T_{pair}$ as a function of $\langle n \rangle$ and $\lambda$ for (a) $U=2t$ and (b) $U=4t$. $T_{pair}$ is always suppressed by SOC for $U=4$ and is dominated by the DOS at Fermi surface for $U=2$.}
\label{Tpair}
\end{figure}

The spin susceptibility is $\chi_{i,j}=-\sum_{k,\omega_{n}} \text{Tr} \{
\sigma_{i} G(k, \omega_{n}) \sigma_{j} G(k,\omega_{n}) - \sigma_{i} F(k,
\omega_{n}) \sigma_{j} F^{\dag}(k,\omega_{n})\}$. $G$ and $F$ can be solved by
Eq.\ref{greenfunction}. Here, we show the result of $\chi_{zz}$.

\begin{eqnarray}
\chi_{zz}&=&-\frac{1}{\beta}\sum_{k,\omega_{n}} \frac{2 (i\omega_{n} + \epsilon_{k})^{2}-2|g_{k}|^{2}+2 \Delta^{2}}{(\omega_{n}^{2}+E_{k,+}^{2})(\omega_{n}^{2}+E_{k,-}^{2})},\nonumber\\
         &=&\sum_{k,\nu}\Big\{ \frac{\tanh(\beta E_{k,\nu}/2)}{2E_{k,\nu}} - \frac{4(\epsilon_{k}^{2}+\Delta^{2})\tanh(\beta E_{k,\nu}/2)}{2E_{k,\nu}(E_{k,\nu}^{2}-E_{k,-\nu}^{2})} \Big\}.
\end{eqnarray}

\begin{figure}
\includegraphics[width=\linewidth]{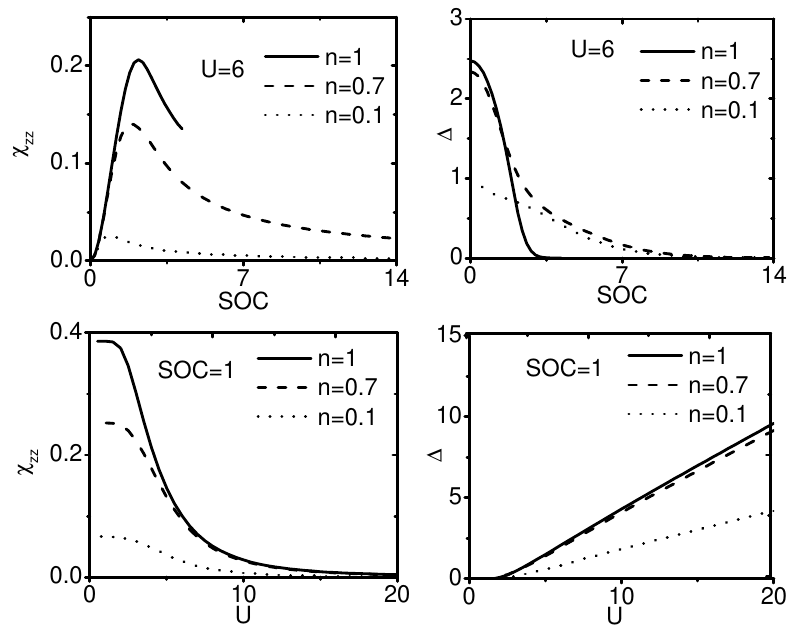}
\caption{$\chi_{zz}$ VS $U$ and $\lambda$ at $T=0$. The behavior of
$\chi_{zz}$}  qualitatively matches with results of DQMC. \label{ss}
\end{figure}

At $T=0$, $\chi_{zz}= \sum_{k}\Big\{
\frac{(E_{k,+}+E_{k,-})^{2}-4(\epsilon_{k}^{2}+\Delta^{2})}{2E_{k,+}E_{k,-}(E_{k,+}+E_{k,-})}\Big\}$.
When $\lambda\ll \{t,\mu,\Delta \}$, we can expand the spin susceptibility by
$\lambda$.
\begin{eqnarray}
\chi_{zz}\doteq \sum_{k}\frac{\Delta^{2}}{E_{k}^{5}}|g_{k}|^{2} \sim \lambda^{2}.
\end{eqnarray}

At zero temperature, spin susceptibility is a quadratic function of $\lambda$.

To investigate the phase fluctuation in mean field framework, we can impose a phase twist on the pairing potential $\Delta_{i}=\Delta e^{i \nabla \theta \cdot \textbf{r}_{j}+ i \partial_{\tau}\theta \cdot \tau}$. The partition function can be expanded by $\partial_{i} \theta$. The partition function has a symmetry to $\theta \rightarrow -\theta$. Therefore, the leading order is $(\partial_{i} \theta)^{2}$.
\begin{eqnarray}
S_{eff}=\frac{1}{2}\int d^{2} \textbf{r} [P (\partial_{\tau} \theta)^2 + D_{s}(\nabla \theta )^2],\nonumber
\end{eqnarray}
with
\begin{eqnarray}
P&=&\sum_{\textbf{k},\nu=\pm}\left\{ \frac{\Delta^{2}}{8 E_{k,\nu}^{3}}\tanh \left(\frac{\beta E_{k,\nu}}{2}\right) + \frac{\varepsilon_{k,\nu}^{2}}{16 E_{k,\nu}^{2}} \text{sech}^{2} \left(\frac{\beta E_{k,\nu}}{2}\right)\right\},\nonumber\\
D_{s}&=&\sum_{\textbf{k},\nu}\left\{ -2t \cos(k_{x})\frac{\varepsilon_{k,\nu}}{E_{k,\nu}}\Big[1-2f(E_{k,\nu})\Big] + 2\lambda \frac{\nu \varepsilon_{k,\nu} \sin^{2} k_{x} }{2E_{k,\nu} \sqrt{\sin^{2} k_{x} + \sin^{2} k_{y}}}\tanh\left(\frac{\beta E_{k,\nu}}{2}\right)\right.\nonumber\\
&&\left.\qquad  - 2 \lambda \nu \frac{\epsilon_{k}^{2} +2 \nu \lambda \epsilon_{k} \sqrt{\sin^{2} k_{x} + \sin^{2} k_{y}} +\Delta^{2}}{2 \epsilon_{k} E_{k,\nu}}\frac{\sin^{2}k_{y}\cos^{2}k_{x}}{(\sin^{2} k_{x} + \sin^{2} k_{y})^{3/2}} \tanh\left(\frac{\beta E_{k,\nu}}{2}\right) \right\}\nonumber\\
&&\qquad + f'(E_{k,\nu}) \sin^{2} k_{x}\left( 2 t + \nu\frac{2 \lambda \cos k_{x}}{\sqrt{\sin^{2} k_{x} + \sin^{2} k_{y}}} \right)^{2},\nonumber
\end{eqnarray}
where $f(x)=(1+e^{\beta x})^{-1}$ is the Fermi-Dirac distribution.

\end{widetext}

\end{document}